\documentclass[twocolumn,showpacs,preprintnumbers,amsmath,amssymb]{revtex4}

\usepackage{graphicx}
\usepackage{amsmath}
\usepackage{color}

\def\ba{\begin{align}}

\begin{document}

\preprint{}

\title{Diffusion Enhances Chirality Selection}

\author{Ryo Shibata} %\thanks{rshibata@rk.phys.keio.ac.jp},
\author{Yukio Saito}  \email{yukio@rk.phys.keio.ac.jp}
\author{Hiroyuki Hyuga} \email{hyuga@rk.phys.keio.ac.jp}

\affiliation{
Department of Physics, Keio University, Yokohama 223-8522, Japan
}

\date{\today}

\begin{abstract}
Diffusion effect on chirality selection in a two-dimensional
reaction-diffusion model
is studied by the Monte Carlo simulation.
The model consists of
achiral reactants A which turn into either of the chiral products, R or S,
in a solvent of chemically inactive vacancies V.
The reaction contains the nonlinear autocatalysis as well as recycling
process, and 
the chiral symmetry breaking is monitored by an
enantiomeric excess $\phi$.

Without dilution a strong nonlinear autocatalysis ensures  chiral
symmetry breaking.
By dilution, the chiral order $\phi$ decreases, and the racemic state
is recovered below the critical concentration $c_c$.
Diffusion effectively enhances the concentration of chiral species,
and $c_c$ decreases as the diffusion coefficient $D$ increases.
The relation between $\phi$ and $c$ for a system with
 a finite $D$
fits rather well to an interpolation formula
between the diffusionless($D=0$) and homogeneous ($D=\infty$) limits.

\end{abstract}

\pacs{05.90.+m, 82.40.Ck, 05.50.+q}% PACS

\maketitle

\section{Introduction}

It has long been known that 
natural biomolecules choose one type of enantiomer 
among the two possible stereostructures,
and the chiral symmetry is broken in life:
Amino acids in proteins are 
all {\small L}-type, 
while sugars in nucleic acids are 
all {\small D}-type \cite{bonner92,kondepudi+98,kondepudi+01}.
There are many studies on the origin of this homochirality,
but the asymmetry caused by various proposed mechanisms
turned out to be very minute
\cite{mason+85,kondepudi+85,meiring87}.
Frank found  more than fifty years ago that a linear autocatalysis 
accompanied by mutual inhibition are necessary in an open system for the
amplification of enantiomeric excess (ee) of one type
 \cite{frank53}.
The model with enantiomeric cross-inhibition is extended to
polymerization models
\cite{sandars03,brandenburg+05}.
Experimental realization of the ee amplification 
is found recently in pyrimidyl alkanol \cite{soai+95},
and explained by a nonlinear autocatalytic process in a closed system
\cite{sato+01,sato+03}.
Addition of a recycling process to this nonlinear system is shown to
realize an asymptotic selection of a unique chiral state
\cite{saito+04a}.
The importance of recycling process is also shown in nonautocatalytic but 
a polymerization-epimerization model \cite{plasson+04}.

All the above arguements assume that the reaction
takes place homogeneously.
For spatially extended systems, however, the inhomogeneity
 affects symmetry breaking in general.
In condensed matter systems, for instance, elements are located densely
in contact with each other,
and interactions propagate through neigboring elements 
so as to cause coherent phase transitions.
In contrast to the equilibrium symmetry breaking in the condensed matter,
organic molecules in the primordial era might have been
distributed dilutely in a solution.
If this were the case, molecular transport is necessary in order 
to bring molecules come into contact to activate autocatalysis.

In a previous letter\cite{saito+04b} we briefly reported
the diffusion effect on the chiral symmetry breaking.
Here we give a more detailed work of the diffusion effect
on the chirality proliferation in an extended system.
After introducing the model and Monte Carlo simulation algorithm
in \S 2,  we consider first the chirality
selection in a diffusionless system, and 
point out some similarity
of the present chemical reaction system to the Ising ferromagnet
in \S 3.
There, the effect of site dilution is also discussed.
In \S 4, the diffusion effect is studied systematically and the 
chiral symmetry breaking is found to be enhanced by the diffusion.
It is qualitatively interpreted that the diffusion effectively increases 
the concentrations of chemically active molecules, and consequently 
promotes autocatalysis.
The idea is summarized in a simple
interpolation formula as for the diffusion coefficient dependence of the 
order parameter and the concentration relation.
The study is summarized in \S 5.

\section{Model and Simulation Algorithms}

The model we study is essentially the same with the one proposed
in a previous letter \cite{saito+04b}.
There are four types of molecules;
an achiral reactant A, two types of product 
enantiomers R and S, and a 
solvent molecule  in a diluted system, which we call
 a vacancy V. 
Molecules are treated as entities which occupy square 
lattice sites and can move randomly thereon.
Double occupancy of a lattice site is forbidden.

%%%%%%%%%%%%%%%%%%%%%
\begin{figure}[h]
\begin{center} 
\includegraphics[width=0.80\linewidth]{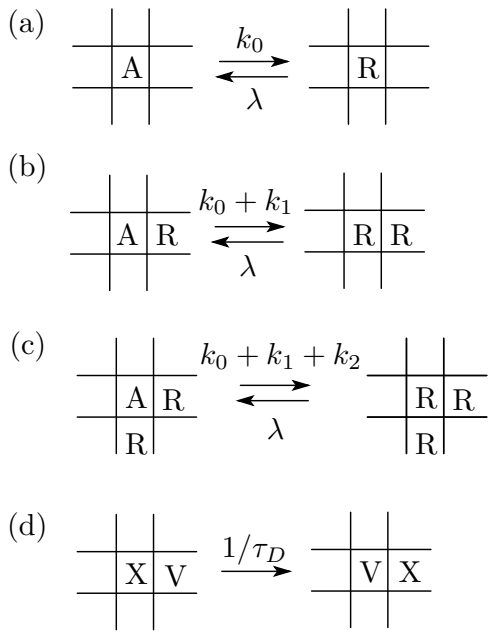} 
\end{center} 
\caption{
Various processes with transition probabilities;
(a) non-autocatalytic, (b) linearly autocatalytic, (c) nonlinearly
autocatalytic reactions with recycling and (d) a vacancy diffusion.
X in the diffusion process (d) represents chemically active species,
A, R or S.
}
\label{fig1}
\end{figure}
%%%%%%%%%%%%%%%%%%%%%

Various processes in the system are summarized in 
Fig.1.
An isolated achiral reactant A turns into a chiral R or S molecule
with a rate $k_0$,
as in Fig.1(a).
When there is one R(S) molecule 
in the nearest neighbor of a reactant A,
the production rate of R(S) increases 
by an additional linear autocatalytic rate $k_1$, 
as in Fig.1(b).
When there are two or more R's(S's) in the nearest neighborhood,
the production rate of R(S) is further enhanced 
by an additional nonlinear autocatalytic rate $k_2$,
as in Fig.1(c).
The rate of back reaction $\lambda$ is assumed constant,
independent of the environment around chiral molecules, R or S
(Fig.1(a)-(c)).
By these chemical reaction processes, the number densities $a,~r$ and $s$
of active molecules A, R and S vary, but
the total number density of active molecules
$a+r+s=c$ is fixed constant, reflecting the assumption 
of a closed system
corresponding to the Soai reaction
 \cite{soai+95}.
The degree of chiral symmetry breaking is represented by the
enantiomeric excess (ee) order parameter
\begin{align}
\phi= \frac{r-s}{r+s} .
\label{eq1}
\end{align}

A vacancy V is inactive as for the chemical reaction, but
plays a vital role in the diffusion process.
When chemically active species, A, R and S, are neighboring to V, 
they exchange their sites with V randomly with a life time $\tau_D$
on average,
as in Fig.1(d).
The diffusion coefficient is then $D=a^2/(4\tau_D)$ with $a$ being the
lattice constant and set to be unity hereafter.
Previously we simulated only the case with $D=1/4$,
because the Metropolis algorithm was used in the Monte Carlo simulation
\cite{saito+04b}.
In the present paper we want a systematic study 
of diffusion effect on the chirality selection, and therefore an
arbitrary variation of the diffusion coefficient $D$ is necessary.
For that purpose, we adopt a waiting time algorithm 
for the Monte Carlo simulation.

The simulation runs as follows:
All the positions which can undertake a process $i$ 
are tabulated in a list;
every site in the list has the same transition probablity $P_i$
to perform the reaction or diffusion process $i$,
shown in Fig.1.
Assume that there are $n_i$ lattice 
sites in the corresponding list of the process $i$. 
Then, the total probability of events in a unit time is obtained as
$P_{\mathrm {tot}}=\sum_{i} P_in_i$.
In other words, in a time interval $dt=1/P_{\mathrm {tot}}$,
one of the process takes place on average. 
By generating a quasi-random number $x$ which is 
uniformly distributed between
0 and $P_{\mathrm {tot}}$, the process $j$ which satisfies
the condition
\begin{align}
\sum_{i=1}^{j-1} P_in_i \le x < \sum_{i=1}^{j} P_in_i 
\label{eq2}
\end{align} 
is to be performed.
The lattice site which performs the process $j$ is listed 
at the $k$-th position in the 
list of 
$n_j$ sites, where
$y=x-\sum_{i=1}^{j-1} P_in_i$ and $k$ is the integer part of 
$y/P_j+1$ as
\begin{align}
k=
\Big[ \frac{y}{P_j}+1 \Big] .
\label{eq3}
\end{align} 
After performing the process $j$ on the site in the $k$-th position
in the list,
the time is advanced by $dt$ and the lists are updated.
We simulate systems with sizes 50$^2$ and 100$^2$, but in the
following we show results obtained for the larger system with a size $100^2$.

\section{Without Diffusion}

From our previous works on a homogeneous system \cite{saito+04a,saito+04b}
the nonlinear autocatalysis with a finte $k_2$ is found essential
for the chiral symmetry breaking, whereas the linear autocatalysis
$k_1$ is inefficient.
Therefore, $k_1$ is set absent hereafter: $k_1=0$.
In order to find out the effect of diffusion,
we consider in this section the developement of chirality in space without
the material transport. 
The effect of diffusion is studied in the next section. Here,
the order of chiral symmetry breaking is studied 
as a function of the strength of the nonlinear autocatalysis or 
of the concentration of active molecular species.

%%%%%%%%%%%%%%%%%%%%%
\begin{figure}[h]
\begin{center} 
\includegraphics[width=0.99\linewidth]{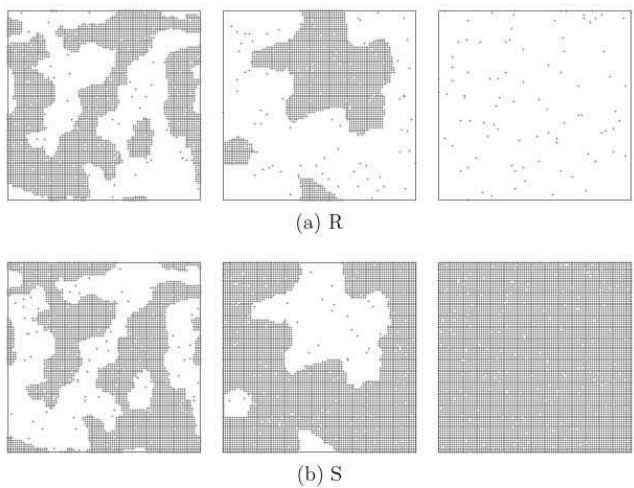} 
%\includegraphics[width=0.32\linewidth]{fig2a1.eps} 
%\includegraphics[width=0.32\linewidth]{fig2a2.eps}
%\includegraphics[width=0.32\linewidth]{fig2a3.eps} \\
%(a) R\\
%\includegraphics[width=0.32\linewidth]{fig2b1.eps}
%\includegraphics[width=0.32\linewidth]{fig2b2.eps}
%\includegraphics[width=0.32\linewidth]{fig2b3.eps}\\
%(b) S
\end{center} 
\caption{Configuration evolution of (a) R and (b) S molecules
in a diffusionless system.
Tmes are $k_0t=50, ~250,~1750$
from left to right.
Parameters are $k_0=\lambda,~k_1=0,~k_2=100k_0$ with $c=1$,
and the system size is $100^2$.
}
\label{fig2}
\end{figure}
%%%%%%%%%%%%%%%%%%%%%
\subsection{Chiral Symmetry Breaking as a function of Nonlinear Autocatalysis}

%%%%%%%%%%%%%%%%%%%%%
\begin{figure}[h]
\begin{center} 
\includegraphics[width=0.99\linewidth]{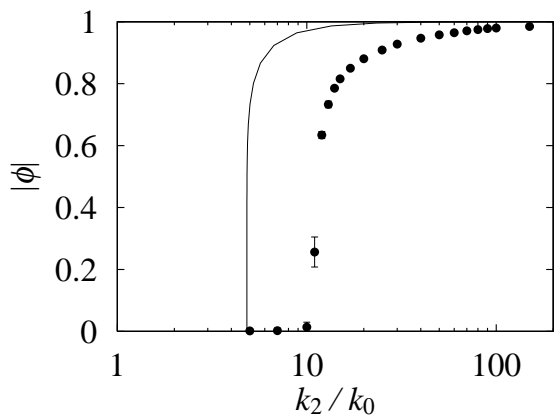}
\end{center} 
\caption{Enantiomeric excess $|\phi|$ versus $k_2/k_0$
at the full occupation $c=1$ in a semi-logarithmic plot.
A  curve represents the temperature dependence
of the magnetization
\cite{yang52} of a square Ising model, where a
temperature is related to the chemical reaction rates as
$k_2/k_0=\exp(4J/k_BT)-1$.
}
\label{fig3}
\end{figure}
%%%%%%%%%%%%%%%%%%%%%

First, we consider the $k_2$-dependence of the chiral symmetry breaking,
when the whole lattice sites are occupied by chemically active species; 
 $c=1$.
When the nonlinear autocatalytic effect is strong enough such as
$k_2= 100k_0$ with  $\lambda=k_0$, achiral reactant A quickly
reduces to chiral species R or S.
Therefore, even by starting from an achiral initial configuration with
only achiral reactants A, domains of R and S are formed 
in the early stage, as shown at a time $k_0t=50$ in Fig.\ref{fig1}.
Then, the domain competition sets in, and those domains surrounded by another
type shrink as shown in the middle column in Fig.2.
Eventually, competition ends up by the dominance of one enantiomeric type, 
S in the case shown in the right column of Fig.2.
In some cases the system decouples into two
inter-penetrating domains of R and S, 
both of which in effect extend to infinity
due to the periodic boundary conditions in the $x$ and $y$ directions:
Then, it takes an enormous time until the minor domain extinguishes 
and the system accomplishes the final equilibrium state.

%%%%%%%%%%%%%%%%%%%%%
\begin{figure}[h]
\begin{center} 
\includegraphics[width=0.90\linewidth]{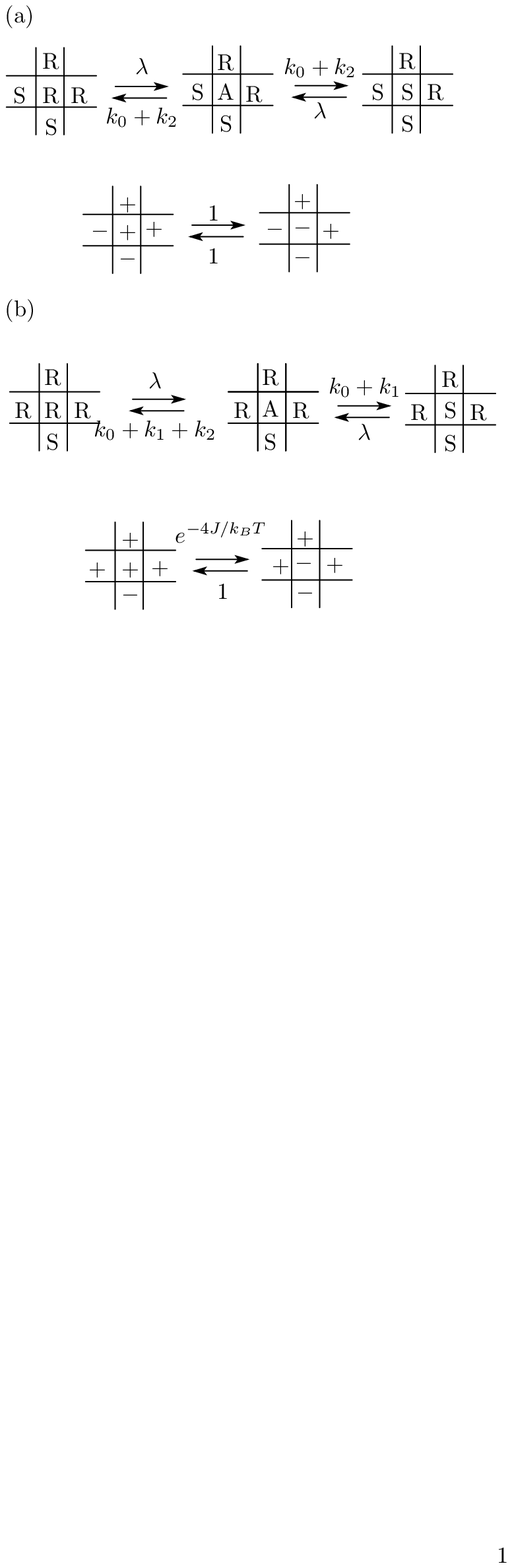} 
\end{center} 
\caption{
Analogies of state transition in the reaction model (above) and 
the Ising ferromagnet (below).
(a) $n=2$, and (b) $n=3$.
}
\label{fig4}
\end{figure}
%%%%%%%%%%%%%%%%%%%%%

In order to obtain the equilibrium value of the order parameter $\phi$
in a reasonable simulation time, 
therefore, we start  simulations
from an initial configuration such that the
whole lattice sites are occupied by R species. 
Equilibrium states are readily realized from 
this initial configuration, and
the equiblirium value of the order parameter $\phi$ thus obtained is
 shown in Fig.\ref{fig3} at various strength 
of nonlinear autocatalysis $k_2/k_0$.
Since the non-autocatalytic reaction $k_0$ drives the system to
the racemic state,
the value of the order parameter $\phi$
decreases as $k_2$ decreases, and vanishes
at a critical value of $k_2/k_0$ around 10:
The system recovers the chiral symmetry, getting into the racemic state.

The qualitative features of domain competition shown in Fig.2 and the
emergence of symmetry breaking at large $k_2$ as shown in Fig.3 
suggest us some resemblance with the phase transition in an Ising ferromagnet.
In fact, one may find some analogy of our reaction system 
to an Ising ferromagnet: 
In the latter, 
each lattice site is occupied by an Ising spin, 
which can be in one of the two states; up or down. 
When a nearest neighboring spin pair have the
same orientation, the energy is low as given by $-J<0$, 
whereare when a pair is in opposite orientations, the energy is high as $J>0$:
The energy difference is set to be $2J>0$.
At a temperature $T$, a spin flips its orientation 
with a transition probability which depends on the associated energy 
change.
Let the spin to flip have $n$ nearest neighboring spins 
with the same orientation. Then,
it will have $4-n$ spins of the same orientation after the flip,
and we denote the transition probability as $W(n \rightarrow 4-n)$.
In order to ensure thermal equilibrium asymptotically,
it is known to be sufficient if
the detailed balance condition between 
the transition probabilities of one process and its reverse 
\begin{align}
\frac{W(n \rightarrow 4-n)}{W(4-n \rightarrow n)}=\exp(-\Delta E/k_BT)
\label{eq4}
\end{align}
is satisfied, where $\Delta E= 4(n-2)J$ represents
an energy change.
For example, when a site is surrounded by two up spins and two down spins,
$n=2$ and the transition probability of an up spin to flip 
downward
is equal to its reverse process, 
as shown in the lower column of Fig.4(a).
Then, in equilibrium the probabilty $P_+(2,2)$ 
of an up spin surrounded by two up spins and two down ones 
is equal to that of a down spin in the same enviornment
$P_-(2,2)$: $P_+(2,2)=P_-(2,2)$.
We associate, for instance, R and S molecules to the up and down Ising spins, 
respectively. Then,
the transitions for $n=2$ in the Ising model 
correspond to transitions between an R and S molecule 
surrounded in the neighborhood
by two R's and two S's ( upper column of Fig.4(a)). 
Since the alternation of R and S is mediated by 
an achiral reactant A, the equilibrium
probabilities of R and S in the configuration
Fig.4(a) are related as
$\lambda P_R(2,2)  = (k_0+k_2) P_A(2,2) =\lambda P_S(2,2)$, 
and thus $P_R(2,2)=P_S(2,2)$, 
consistent with our assignment of R and S molecules to up and down spins.
When a site is surrounded by three R's and one S molecule 
as in the upper column of Fig.4(b),
the equilibrium probability that the site is occupied by an R molecule 
$P_R(3,1)$ is higher than the one
by an S molecule $P_S(3,1)$. They are related as 
$\lambda P_R(3,1)  =(k_0+k_2) P_A(3,1) $ and $  k_0 P_A(3,1)=\lambda P_S(3,1)$,
 and thus $P_R(3,1)=(1+k_2/k_0) P_S(3,1)$: 
 At a lattice site surrounded by three R's and an S,
the probability of finding an R molecule 
is higher than that of finding an S molecules by a 
factor $(1+k_2/k_0)$.
 For the Ising system, the corresponding situation is the spin on the site
surrounded by three up's and one down; $n=3$.
The relation between probabilities that a site is
occupied by an up spin and by a down is
$e^{-4J/k_BT} P_+(3,1)=P_-(3,1)$.
Thus, we may assume that the ratio $(1+k_2/k_0)$ corresponds 
in the  Ising model to the ratio of the 
transition probabilities $\exp(4J/k_BT)$.
(See the lower column of Fig.4(b)).
In other word, the nonlinear autocatalytic rate $k_2$ in our chiral model might
correspond to the coupling constant of the two-dimensional Ising model as
\begin{align}
k_2/k_0=e^{4J/k_BT}-1 .
\label{eq5}
\end{align}
Since the numbers of R and S molecules correspond to those of
up and down spins in the Ising ferromagnet,
the chiral order parameter $\phi$ corresponds to the 
spontaneous magnetization $M$ of the Ising model.
The latter is
exactly known to behave as \cite{yang52}
$M=[1-\sinh(2J/k_BT)^{-4}]^{1/8}$.
By using the relation (\ref{eq5}), the magnetization $M$ is plotted in
Fig.3 by a continuous curve.

The transition temperature $T_c$ of the Ising model on the square
lattice is exactly known to be at $k_BT_c/2J=1/\ln(1+\sqrt{2}) \approx 1.135$.
Thus in the present reaction model
the chiral symmetry breaking is expected to take place
at $k_2 \ge k_{2c}=2( \sqrt{2}+1) k_0 \approx 4.83k_0$.
However,
the order parameter $\phi$ obtained by simulations decreases faster 
than the Ising magnetization as $k_2$ decreases:
For the chiral symmetry breaking a larger value of the nonlinear autocatalysis
$k_2 \approx 10 k_0$ is required, as shown in Fig.\ref{fig3}.
This is because  of
the approximate character of the relation between the Ising 
coupling constant $J$ and the nonlinear autocatalytic rate $k_2$.
The production rate in our model 
is independent of the number of neighboring R(S) species 
as far as it is more than 2, whereas transition rate to the up spin
state in an Ising model increases as the
neighboring number of up spins increases:
This means that our coupling constant $k_2$ is effectively
weaker than the Ising coupling parameter.
Also the existence of an achiral species is not included in the Ising model.
The decomposition process from a chiral species R or S 
back to an achiral reactant A induces fluctuation which breaks ordering,
and a chirally ordered state in our reaction model 
is less stable than the ferromagnetic state in the Ising ferromagnet:
The strong nonlinearity is necessary for the symmetry breaking.
In fact, the critical value for the chiral symmetry breaking $k_{2c}$ 
depends on
the decomposition rate $\lambda$, though the $k_2$ versus $J$ relation in
eq.(5) is independent of $\lambda$.
From simulations 
with a large  $\lambda=10k_0$, the order is 
found to be
easily broken at 
$k_{2c} \approx 23k_0$,
whereas for small $\lambda=0.1k_0$ the order holds until
$k_{2c} \approx 9k_0$.

\subsection{Dilution Effect on Chiral Symmetry Breaking}

We now keep the nonlinear autocatalitic effect as strong as $k_2/k_0=100$,
and 
consider the dilution effect ($c <1$) on the chiral symmetry breaking.
If there is no material transport, 
the behavior of the system depends strongly on the arrangement of
vacant sites. Therefore, 
in order to obtain the order parameter $\phi$
we average 
9 samples at concentrations lower than 
$c=0.7$, whereas 3 samples seem sufficient at higher concentrations.
As the system is diluted,
the probability that active chemical species
are contiguous to each other decreases.
In the present reaction model, however,
a reactant A should be surrounded
by chiral products R or S to sustain the autocatalysis.
Therefore, the chiral symmetry breaking is possible only
 at concentrations at least larger than the site percolation threshold 
 of the square lattice 
 $c_p=0.592746$
 \cite{stauffer85,ziff92}.
Below $c_p$, the information of one site is unable to propagate through
the whole system, and thus the system cannot be in a single state.
The order parameter $\phi$ actually decreases as the system is diluted,
as shown in Fig.5a,  and the system becomes racemic
at concentrations lower than the critical value $c_c(D=0)$.
By assuming a critical concentration at $c_c(D=0) \approx 0.639$,
the order parameter $\phi$ shows a power-law dependence 
$\phi \propto \sqrt{c-c_c(D=0)}$, as shown in Fig.5(b).
The critical concentration  $c_c(D=0)$ without diffusion 
is close to but larger than $c_p$.

A rather large discrepancy between $c_c(D=0)$ and $c_p$ 
might be understood by using the similarity of
the present diluted reaction model to a site-diluted Ising ferromagnet,
where lattice sites are diluted substitutionally by nonmagnetic impurities
\cite{stinchcombe83}.
In a site-diluted Ising model, the transition temperature is known to decrease
as the concentration of nonmagnetic impurity increases,
and the phase transition vanishes 
at the critical concentration close to the percolation threshold $c_p$
\cite{neda94}:
The transition temperature is zero at the critical concentration.
If the relation (5) holds, however,
our reaction model with $k_2=100k_0$ corresponds to an Ising ferromagnet
at a finite temperature $k_BT/2J=0.433$ or $T/T_c=0.38$.
A thermal fluctuation favors the recoveray of the symmetry 
even if a percolating cluster exists, and
this may be the reason that the critical concentration $c_c(D=0)$ 
is larger than the percolation threshold $c_p$.

%%%%%%%%%%%%%%%%%%%%%
\begin{figure}[h]
\begin{center} 
\includegraphics[width=0.75\linewidth]{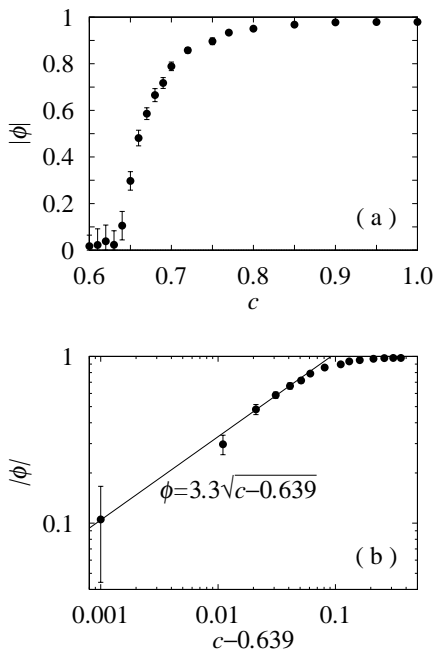}
\end{center} 
\caption{Enantiomeric excess $|\phi|$ versus $c$
in a diffusuioness system
with a strong nonlinear autocatalysis $k_2/k_0=100$.
}
\label{fig5}
\end{figure}
%%%%%%%%%%%%%%%%%%%%%

\section{Diffusion Enhancement of the Chirality Selection}

We now study the diffusion effect on the chirality selection
in a diluted system.
The transport of material in general increases the chance 
for an achiral reactant A to meet chiral products R and S
in their lifetimes,  and may enhance the autocatalytic effect.
When the diffusion is extremely fast, the system becomes homogeneous:
This case will be studied in the following subsection analytically.
With a finite diffusion coefficient, 
numerical simulation is the only possible method to study, 
and the results so obtained will be 
analyzed by the interpolation formula in the second subsection.

\subsection{Homogeneous Case with $D=\infty$}

When the diffusion is extremely fast, the system becomes 
homogeneous.
Then, the simple mean field approximation leads to the 
rate equations for the homogeneous concentrations;
\begin{align}
\dot r = k(r)a - \lambda r
\nonumber \\
\dot s = k(s)a - \lambda s ,
\label{eq6}
\end{align}
with the rate constant $k(r)=k_0+4k_1 r + 6k_2r^2$. 
Here in this subsection we restore the rate $k_1$ of
the linear autocatalytic reaction, in order to be general.
The factors $4r$ and $6r^2$ represent the probabilities 
that at least one or two of the
four neighboring sites are occupied by R species, respectively,
in the sense of mean field approximation.
Since the system is assumed to be closed,
the total concentration of chemically active species
satisfies the concervation law; $a+r+s=c=$const.
Then, the total concentration of chiral products $u=r+s$ is shown to varies as
\begin{align}
\dot u= [2k_0+4k_1u+3k_2u^2(1+\phi^2)](c-u)
-\lambda u ,
\label{eq8}
\end{align}
whereas the order parameter $\phi$ evolves as
\begin{align}
\dot \phi= \frac{c-u}{u} \phi [-2k_0+3k_2u^2(1-\phi^2)] .
\label{eq7}
\end{align}
Since $0 \le u \le c$, eq.(8) is similar to the time-dependent
Landau equation with nontrivial fixed points 
if $u^2 > 2k_0/(3k_2)$.
The other point to remark is that eq.(8) does not depend explicitly 
on the linear autocatalytic rate $k_1$;
 $\phi$ depends on $k_1$  only implicitly through $u$.
 Thus the linear autocatalysis alone cannot break the chiral symmetry.

The equilibrium values of $u$ and $\phi$ are calculated from the
algebraic equations given by $\dot \phi=\dot u=0$.
The spontaneous chiral symmetry breaking takes place at
a critical concentration $c_c(D={\infty})$ when 
the diffusion is infinitely fast:
The order parameter $\phi$ of a homogeneous system takes a finite value
for $c_c(D={\infty})<c \le 1$.
For $k_1=0$  the order parameter
$\phi$ is explicitly written as a function of the concentration $c$ as
\begin{align}
\phi^2=1-\frac{6k_0k_2}{\lambda^2}
\left( c-\sqrt{c^2-\frac{2\lambda}{3k_2}} \right)^2,
\label{eq9}
\end{align}
and the critical concentration is calculated to be 
$c_c(D={\infty})=(4+\lambda/k_0)/\sqrt{k_0/24k_2}$.
It takes the value $c_c(D={\infty})=0.1020$
for the parameter values $k_1=0,~k_2=100k_0,~\lambda=k_0$.

\subsection{Finite Diffusion Constant}

%%%%%%%%%%%%%%%%%%%%%
\begin{figure}[h]
\begin{center} 
\includegraphics[width=0.8\linewidth]{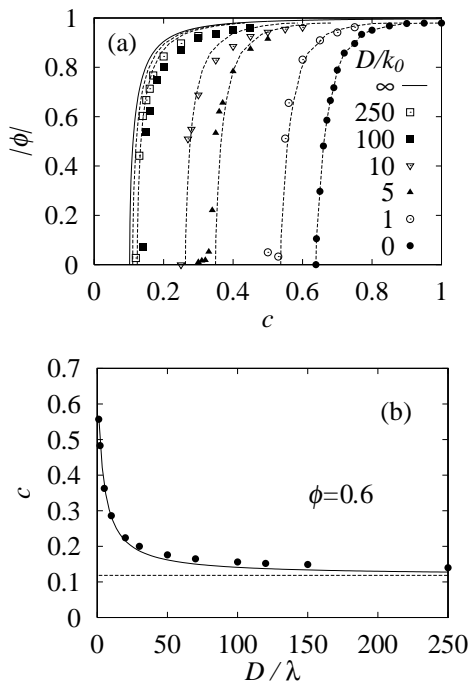}
\end{center} 
\caption{(a) Enantiomeric excess $|\phi|$ versus $c$
at various values of the diffusion coefficient $D/k_0$.
The reaction parameters are $k_2/k_0=100,~k_1=0$ and $\lambda=k_0$.
A continuous curve represents the relation (9)  for a homogeneous system.
Dashed curves represent interpolation (10) at various diffusion coefficients.
(b) Diffusion coefficient dependence of the concentration $c$ where
the order parameter takes the value $\phi=0.6$.
A curve represents the interpolation (10) with a parameter $A=0.235$.
The dotted line represents the homogeneous case with $c(0.6; D=\infty)=0.118$.
}
\label{fig6}
\end{figure}
%%%%%%%%%%%%%%%%%%%%%

The emergence of chiral order at a finite vacancy diffusion is studied
 by the Monte Carlo simulation. 
It is found that the symmetry breaking takes place
rather rapidly even if we start from a completely achiral
initial configuration with all the chemically active molecules being 
achiral reactant A as $a=c$:
The order parameter approaches quickly the asymptotic value, 
though which enantiomer
dominates over the other is determined randomly in each simulation run.
Therefore, we initiated the simulation from the complete achiral state
with only A and V species.

The steady-state value of ee or $|\phi|$ at various values of 
diffusion coefficient $D$ is shown as a function of the 
concentration of active species $c$ by symbols in Fig.\ref{fig6}(a).
A continuous curve represents the relation (9)  for a homogeneous system
with $D=\infty$.
The most striking feature is that  $\phi - c$ 
relations at different $D$'s look similar but with shifts
to the lower concentration  as $D$ increases.
The $D$-dependence of the shift might be considered in the following way.
The chiral products R or S once produced
diffuse around in a diffusion area of about $D/\lambda$ sites during the
life time $\lambda$. 
Therefore, 
in a system where diffusion is allowed
the effective density of chiral catalysts 
is enhanced by a factor $1+A D/\lambda$ 
in comparison to the
system without diffusion.
A proportionality constant $A$ will be determined later
by fitting to the simulation results.
With a very large diffusion $D$, however, there will be overlappings
of various diffusion areas, and the diffusive enhancement may saturate.
Thus, even with an infinitely fast diffusion, a finite concentration of active 
species is necessary, as is given by eq.(9).

The diffusive enhancement of the effective concentration
means that in order to reach a fixed value
of the order parameter $\phi$,
the required concentration of active species $c(\phi; D)$ for
a system with a finite diffusion coefficient $D$
can be less than  $c( \phi; D=0)$ for
a diffusionless system.
Here we propose a simple formula for $c(\phi; D)$
interpolating between the diffusionless limit $c(\phi ; D=0)$
and the homogeneous limit $c(\phi ; D=\infty)$ as
\begin{align}
c(\phi ;D)=
 \frac{c( \phi; 0) + c(\phi; \infty) ( A D/\lambda)}{1+(A D/\lambda)},
\label{eq10}
\end{align}
where $c(\phi; 0)$ is obtained by the simulation (Fig.5(a)), whereas 
$c(\phi; \infty)$ is obtained by solving eq.(9) as
\begin{align}
c(\phi; {\infty}) = \frac{1}{\sqrt{24k_0 k_2}} \Big( \lambda \sqrt{1-\phi^2}+
\frac{4k_0}{\sqrt{1-\phi^2}} \Big) .
\label{eq11}
\end{align}
To determine the constant $A$, we fit the formula (10) with 
simulation data at $\phi=0.6$. 
The reason for the choice of a non-zero value of $\phi$ is
that the value of $c(\phi; D)$ is expected to be more accurate
away from the transition point, $\phi=0$.
The fit gives the value $A \approx 0.235$, and agrees
well at small $D$ but deviates systematically at large $D$,
as shown in Fig.\ref{fig6}(b).

With this value of $A$, we interpolate the whole $\phi-c$
curve for arbitrary $D$ by eq.(10), as is drawn by broken curves in Fig.6(a).
The trend of the down shifting in the concentration as the diffusion
increases is well reproduced,
even though the fitting is not perfect, especially at 
a large diffusion coefficient $D$.
There, we have so far no plausible theory
on the $D$-dependence of $c(\phi; D)$.

\section{Summary and Discussions}

We study diffusion effect on the chirality selection 
during a chemical production of chiral molecules, R or S, 
from an achiral substrate A
 in a solution by means of a square lattice model.
In addition to the usual reactions of the production and the decompsition of
chiral species, autocatalytic processes are incorporated such that 
an achiral substrate surrounded by many enantiomers of one type
increases the turn-over rate to that type.
Specifically, the nonlinear autocatalysis which takes place
when the substrate is surrounded by two or more enantiomers of one type
is found to be essential in the chiral symmetry breaking.

When all the lattice sites are ocuupied by chemically active species,
A, R or S, 
the spontaneous symmetry breaking has some similarity to the
phase transition of an Ising ferromagnet. The strength of the
nonlinear autocatalysis $k_2$ in our chemical model
is approximately related to the Ising coupling constant.
The chemical model at very large $k_2$
evolves through the domain competition,
similar to the dynamics of the Ising model at low temperatures.
As $k_2$ decreases, the chiral order extinguishes, but
earlier than the expectation from the Ising model.
The discrepancy may be due to the larger degrees of freedom;
three in the chemical model whereas two in the Ising model.

Inclusion of vacancies or dilution
suppresses the degree of the chiral symmetry breaking
or the enantiomeric excess  $\phi$, 
and below the critical concentration of chemically active species, the system 
is in a racemic state with a chiral symmetry. 
In a diffusionless system ($D=0$) with a strong nonautocatalysis, 
the system is similar to the site-diluted Ising model, and
the critical concentration is close to but larger than
the percolation threshold $c_p$.

Diffusion enhances the chiral order $\phi$, 
since the migration of chiral products enhances the chance of
molecular collision and the autocatalytic process as well.
In the limit of an infinitely fast diffusion ($D=\infty$), 
the system behaves as if it is homogeneous and 
 the rate equation in the mean field approximation
 describes the chiral symmetry breaking:
The steady state solution is obtained analytically in this limit.
At a finite diffusion coefficient $D$, the concentration dependence of $\phi$ 
is shifted from the diffusionless limit ($D=0$) to the homogeneous
limit ($D=\infty$).
The behavior is interpreted 
as the diffusive enhancement of concentrations of
chiral species, and is summarized in an interpolation formula 
of $c$-dependence of $\phi$ or $c(\phi; D)$ between the two limits.

%%%%%%%%%%%%%%%%%%%%%%%%%%%%%%%%%%%%%%%%%%%%%%%%%%%%%%%5


\begin{thebibliography}{99}

\bibitem{bonner92}
W.A. Bonner, Origins of Life {\bf 21}, 407 (1992).

\bibitem{kondepudi+98}
D. Kondepudi and I. Prigogine, {\it Modern Thermodynamics},
(John Wiley, Chichester, 1998).

\bibitem{kondepudi+01}
D. K. Kondepudi and K. Asakura, Acc. Chem. Res. {\bf 34}, 946 (2001).

\bibitem{mason+85}
S. F. Mason and G. E. Tranter, Proc. R. Soc. Lond. {\bf A 397}, 45 (1985).

\bibitem{kondepudi+85}
D. K. Kondepudi and G. W. Nelson, Nature {\bf 314}, 438 (1985).

\bibitem{meiring87}
W. J. Meiring, Nature {\bf 329}, 712 (1987).

\bibitem{frank53}
F. C. Frank, Biochim. Biophys. Acta {\bf 11}, 459 (1953).

\bibitem{sandars03}
P.G.H. Sandars, Orig. Life Evol. Biosph. {\bf 33}, 575 (2003). 

\bibitem{brandenburg+05}
A. Brandenburg, A. Andersen, M. Nilsson and S. H\"ofner, 
Orig. Life Evol. Biosph. {\bf 35}, 225 (2005).

\bibitem{soai+95}
K. Soai, T. Shibata, H. Morioka and K. Choji, Nature {\bf 378}, 767 (1995).

\bibitem{sato+01}
I. Sato, D. Omiya, K. Tsukiyama, Y. Ogi and K. Soai, 
Tetrahedron Asymmetry {\bf 12}, 1965 (2001).


\bibitem{sato+03}
I. Sato, D. Omiya, H. Igarashi, K. Kato, Y. Ogi, K. Tsukiyama and K. Soai, 
Tetrahedron Asymmetry {\bf 14}, 975 (2003).

\bibitem{saito+04a}
Y. Saito and H. Hyuga, J. Phys. Soc. Jpn {\bf 73}, 33 (2004).


\bibitem{plasson+04}
R. Plasson, H. Bersini, and A. Commeyras, Proc. Natl. Acad. Sci.
{\bf 101}, 16733 (2004).

\bibitem{saito+04b}
Y. Saito and H. Hyuga, J. Phys. Soc. Jpn {\bf 73}, 1685 (2004).

\bibitem{stauffer85}
D. Stauffer, {\it Introduction to Percolation Theory} 
(Taylor and Francis, London, 1985).


\bibitem{ziff92}
R.M. Ziff,  Phys. Rev. Lett. {\bf 69}, 2670 (1992).


\bibitem{stinchcombe83}
R.B. Stinchcombe, in {\it Phase Transition and Critical Phenomena}, 
edited by C. Domb and J.L. Lebowitz 
(Academic, New York, 1983), Vol.7, p.151.


\bibitem{neda94}
Z. N\'eda, J. Phys. I France {\bf 4}, 175 (1994).


\bibitem{yang52}
C.N. Yang, Phys. Rev. {\bf 85}, 808 (1952).

\end{thebibliography}
\end{document}